\documentclass[11pt]{article}
\usepackage{latexsym}
\usepackage{amssymb}
\usepackage{epsf}
\usepackage{amsmath}
\usepackage{slashed}
\usepackage[dvipdfm]{graphicx}
\usepackage{color}
\usepackage{ifpdf}

\newcommand{\Tr}{\rm Tr}
\newcommand{\ro}[1]{\sqrt{\mathstrut #1}}
\newcommand{\lef}{\left( \hspace{-0.5ex}}
\newcommand{\rig}{\hspace{-0.5ex} \right)}

\newcommand{\bea}{\begin{eqnarray}}
\newcommand{\eea}{\end{eqnarray}}
\newcommand{\bear}{\begin{array}}
\newcommand {\eear}{\end{array}}
\newcommand{\bef}{\begin{figure}}
\newcommand {\eef}{\end{figure}}
\newcommand{\bec}{\begin{center}}
\newcommand {\eec}{\end{center}}

\def\GEV#1{10^{#1}{\rm\,GeV}}

\def\lrf#1#2{ \left(\frac{#1}{#2}\right)}
\def\lrfp#1#2#3{ \left(\frac{#1}{#2} \right)^{#3}}

\begin{document}
\pagestyle{empty}

\begin{flushright}
TU-918
\end{flushright}

\vspace{3cm}

\begin{center}

{\bf\LARGE Cosmologically viable gauge mediation}
\\

\vspace*{1.5cm}
{\large 
Hiraku Fukushima, Ryuichiro Kitano and Fuminobu Takahashi
} \\
\vspace*{0.5cm}

{\it Department of Physics, Tohoku University, Sendai 980-8578, Japan}\\
\vspace*{0.5cm}

\end{center}

\vspace*{1.0cm}

\begin{abstract}
{\normalsize 
Gauge mediation provides us with a complete picture of supersymmetry
breaking and its mediation within the effective field theories, and thus
allows us to discuss consistencies with low-energy particle physics as
well as cosmological observations.
We study in detail the cosmological evolution of the pseudo-modulus field
 in the supersymmetry breaking sector and also the production of the
 gravitinos in the early Universe in a simple (but a complete) model of
 gauge mediation.
%
%
Under fairly reasonable assumptions, it is found that there exists a
 parameter region where dark matter of the Universe is explained by
both thermally and non-thermally produced gravitinos, while the baryon asymmetry of the
 Universe is generated through the thermal leptogenesis.
}
\end{abstract} 

\newpage
\baselineskip=18pt
\setcounter{page}{2}
\pagestyle{plain}
\baselineskip=18pt
\pagestyle{plain}

\setcounter{footnote}{0}

\tableofcontents

\newpage 

\section{Introduction}
\label{sec:1}

If supersymmetry is realized in nature, there should be a fermionic
superpartner of the graviton, the gravitino. Since supersymmetry (SUSY) must be
spontaneously broken, the gravitino is massive. Nevertheless, it becomes
massless when we send the Planck scale to infinity, and thus the
gravitino is the lightest superparticle under an assumption that the
gravity is the weakest force to talk to the SUSY breaking
sector.
This discussion already motivates us to consider the gravitino as dark
matter of the Universe, regardless of the hierarchy problem. 
The assumption that the gravity as the weakest communication is always
true in gauge mediation scenarios, where various explicit models are
known. We take one of the simple models of gauge mediation as an
example, and demonstrate how such a scenario is consistent with
observations of the Universe.

In linearly realized models of SUSY breaking, i.e., in the
O'Raifeartaigh models, there is a chiral multiplet $S$ whose
$F$-component acquires a vacuum expectation value (VEV). The fermionic
component of $S$ is the Nambu-Goldstone fermion (the goldstino), and it
is eaten by the gravitino in supergravity theory. The $S$ multiplet also
contains a complex (or two real) physical scalar field which radiatively
obtains a mass.
Since the scalar component has no potential at tree level or at the
supersymmetric level, it is called the pseudo-modulus, that may have 
cosmological importance. We will consider the dynamics of this field
later in detail.

In the early Universe, the gravitinos are produced in various ways. The
scattering of particles with superparticles in the thermal plasma, for
example, produce
gravitinos~\cite{Moroi:1993mb,Kawasaki:1994af,Moroi:1995fs,de
Gouvea:1997tn,Bolz:1998ek,Bolz:2000fu,Pradler:2006qh,Pradler:2006hh,
Rychkov:2007uq}. The fraction of the gravitino energy density,
$\Omega_{3/2}$, from such thermal productions is proportional to the
reheating temperature of the Universe, $T_R$.\footnote{
Throughout this paper, the reheating refers to the decay of the inflaton.
}
Since $\Omega_{3/2}$ is bounded by the total dark matter density
$\Omega_{\rm DM} \sim 0.2$, $T_R$ needs to be low enough, and that
possibly causes a tension with scenarios of baryogenesis. For example,
the thermal leptogenesis~\cite{Fukugita:1986hr} requires $T_R \gtrsim 10^9$~GeV~\cite{Buchmuller:2005eh} which is too
high in typical gauge mediation models.
Gravitinos can also be produced non-thermally such as by the decays of
the string
moduli~\cite{Hashimoto:1998mu,Endo:2006zj,Nakamura:2006uc,Dine:2006ii,Endo:2006tf},
the inflaton~\cite{Kawasaki:2006gs,Asaka:2006bv,
Endo:2007ih,Endo:2007sz,Takahashi:2007tz} and/or the pseudo-modulus
field~\cite{Dine:1983ys,Coughlan:1984yk,Banks:1993en,Joichi:1994ce,Ibe:2006rc,Hamaguchi:2009hy,Kamada:2011ec}.

Once we set up a whole picture of a gauge mediation model, the gravitino
abundance through the thermal scattering processes and also through
non-thermal productions via the pseudo-moduli decays are calculable.
For example, in Ref.~\cite{Ibe:2006rc,Hamaguchi:2009hy}, a simple model
of gauge mediation~\cite{Kitano:2006wz} has been considered, and indeed
it has been found that the non-thermally generated gravitino can explain
the dark matter abundance in a theoretically motivated parameter region.

In fact, the dynamics of the pseudo-modulus is somewhat complicated. As
the theorem of Ref.~\cite{Komargodski:2009jf} states, the O'Raifeartaigh-type model generally has
multiple potential minima, and the one which gives the gaugino mass
correctly is never the lowest one.
Although the theorem does not apply to the model of
\cite{Kitano:2006wz}, the situation is similar due to the existence of a
supersymmetric vacuum where the messenger fields condense.
An important assumption made in the studies of
Refs.~\cite{Ibe:2006rc,Hamaguchi:2009hy} is that the field value of the
pseudo-modulus after inflation is located far away from its origin so
that it never approaches to the true minimum.
Under this assumption, it has been found that the pseudo-modulus
successfully  settles down at the meta-stable SUSY breaking vacuum.

However, it has been pointed out recently that the above assumption is
not necessary when we take into account the thermal effects on the
potential~\cite{Dalianis:2010yk,Dalianis:2010pq}.
Once we start from the origin of the pseudo-modulus, the messenger fields
are massless and thus they are thermalized. The thermal effects of the
messenger fields contribute to the potential of the pseudo-modulus. A
parameter region where the meta-stable minimum is selected along the
thermal history was identified.

It is then natural to ask whether the gravitino abundance is consistent with observations
in such a scenario. Since the messenger fields are abundant in the
thermal plasma, it looks dangerous for the production of the goldstino
component of the gravitino which is directly coupled to the messenger
fields.
In order to see if this scenario is viable, one needs to follow the
cosmological history and calculate the gravitino abundance both from the
thermal and non-thermal processes.

In this paper, we explicitly calculate the gravitino abundance in a
scenario where the initial position of the pseudo-modulus is close to the origin
where the thermal potential is minimized.
We find various non-trivial behavior of the pseudo-modulus depending on
model parameters.
By numerically following the motion of the pseudo-modulus in the field
space, it is found that the coherent oscillation of the pseudo-modulus
eventually dominates the energy density of the Universe in a wide range of the parameter space 
and there the non-thermal production of the gravitinos by its decay can explain the
right amount of dark matter. Moreover, it is found that the reheating
temperature after inflation can be much higher than the one required by
the thermal leptogenesis scenario, without having a trouble with the
overproduction of the gravitinos.
By considering the dilution of the baryon asymmetry by the entropy
production from the decays of the pseudo-modulus, the required reheating
temperature is higher or comparable to $10^{12}$~GeV, with which one can
explain both dark matter and baryon asymmetry of the Universe.

The paper is organized as follows.  In section 2 we set up a gauge
mediation model which we use for the study of cosmology.
The dynamics of the pseudo-modulus is studied in section 3 and the
results are used in section 4 for the calculation of the gravitino
abundance generated from scattering processes in the thermal bath.
The non-thermal component is calculated in section 5.

\section{A model of gauge mediation}
\label{sec:2}
We study the low-energy effective theory of O'Raifeartaigh type SUSY
breaking model coupled with the messenger fields.  After integrating out
the massive fields, the K$\ddot{\rm a}$hler potential and the
superpotential are written as
\begin{align}
&K= f^{\dagger} f + \bar{f}^{\dagger} \bar{f} + S^{\dagger} S - \frac{(S^{\dagger} S)^2}{\Lambda^2} + \cdots , 
\label{kaler} \\
&W=m^2 S - \lambda S f \bar{f} + c
\label{super} ,
\end{align}
where $S$ is a gauge singlet field and $f$ and $\bar f$ are the messenger fields
which carry standard model quantum numbers.
We take the messenger fields $f$ and $\bar f$ to transform as $\bf 5$ and $\bf \bar{5}$ under SU(5),
and the messenger number $N=1$ for simplicity.
$\Lambda$ is the cutoff scale of the effective theory,
which is typically the mass scale of the massive fields of the O'Raifeartaigh models. 
 The U(1)$_R$ charge assignment is $R(S) = 2$ and $R(f {\bar f}) = 0$, whereas 
the constant term $c$ represents the R-symmetry breaking supergravity effect and contributes to 
the cosmological constant. We take $m$, $\lambda$, $c$ and $\Lambda$ real and positive by appropriate 
redefinition of the fields and the $U(1)_R$ transformation.
There is a quantum correction to the K$\ddot{\rm a}$hler potential from the interaction term
$\lambda S f {\bar{f}}$,
\begin{align}
K_{\rm 1-loop} = - \frac{5 \lambda^2}{( 4 \pi )^2} S^{\dagger} S \log \frac{S^{\dagger} S}{\Lambda^2},
\label{kaler1}
\end{align}
at the one-loop level. As we shall see, if this radiative correction is too large, the SUSY breaking
vacuum is destabilized, and so, $\lambda$ is bounded above.

Once we take into account the supergravity effects,
a SUSY breaking vacuum appears~\cite{Kitano:2006wz},
\begin{align}
\langle S \rangle = \frac{\ro 3}{6} \frac{\Lambda^2}{M_{\rm pl}}, \ \  
\langle f \rangle = \langle \bar{f} \rangle = 0, 
\label{SUSYbmin}
\end{align}
where we have neglected the radiative correction to the K\"ahler potential. 
Here $M_{\rm pl} \simeq 2.4 \times 10^{18}$\,GeV is the reduced Planck scale. 
The $F$-term of $S$ is
\begin{align}
F_S = m^2,
\end{align}
in this vacuum, and therefore  SUSY is indeed broken. The messenger fields acquire a SUSY 
mass $M_{\rm mess} = \lambda \langle S \rangle$
 through the interaction with $S$.
Note that $S$ develops a VEV along the real component, while the imaginary component vanishes
in this vacuum. 
The constant term $c$ is fixed so that the cosmological constant is cancelled at the SUSY breaking
vacuum, 
\begin{align}
c \;\simeq\; m_{3/2} M_{\rm pl}^2 ,
\end{align}
where the gravitino mass is given by
\begin{align}
m_{3/2} = \frac{m^2}{\ro{3} M_{\rm pl}} .
\label{gravitinomass}
\end{align}

So far we have neglected the radiative correction (\ref{kaler1}), which could destabilize the SUSY breaking 
vacuum. In order to see its effect, let us study the mass terms of the $S$ field and the messenger fields
(we take $\bar{f} = f$), including the correction:
\begin{gather}
V_{\rm mass}
=
(S^{\dagger} \ S) {\cal M}_S^2 \lef \begin{array}{@{\,}c@{\,}} S \\ S^{\dagger} \end{array} \rig + 
(f^{\dagger} \ f) {\cal M}_f^2 \lef \begin{array}{@{\,}c@{\,}} f \\ f^{\dagger} \end{array} \rig , \\
{\cal M}_S^2\; \simeq\; \frac{m^4}{\Lambda^2} \lef 
\begin{array}{cc}
4 & -\frac{15 \lambda^2}{8 \pi^2} \frac{M_{pl}^2}{\Lambda^2} \\
-\frac{15 \lambda^2}{8 \pi^2} \frac{M_{pl}^2}{\Lambda^2} & 4
\end{array} \rig \ \ , \ \ \\
 {\cal M}_f^2 \;=\; \lef 
\begin{array}{cc}
 \lambda^2 \langle S \rangle^2 & - \lambda m^2 \\
 - \lambda m^2 & \lambda^2 \langle S \rangle^2
\end{array} \rig,
\end{gather}
where  we have fixed the SUSY breaking vacuum as (\ref{SUSYbmin}) for simplicity,
and we have dropped terms proportional to $\lambda^2$ in the diagonal components of ${\cal M}_S^2$.
In order for the SUSY breaking minimum to be (meta)stable,
$\det m_S^2 > 0$ and $\det m_f^2 > 0$ must be met, namely,
\begin{align}
\frac{12 m^2 M_{\rm pl}^2}{\Lambda^4} \;<\;\lambda \; < \; \frac{4 \sqrt{2} \pi}{\sqrt{15}} \frac{\Lambda}{M_{\rm pl}}.
\end{align}
Note that $\lambda$ is  bounded both above and below.

There is also a SUSY preserving vacuum at
\begin{align}
\langle S \rangle_{\rm SUSY} = 0, \ \ \langle f \rangle_{\rm SUSY} = \langle \bar{f} \rangle_{\rm SUSY} = 
\ro{\frac{m^2}{\lambda}},
\end{align}
and the messenger directions are tachyonic for 
\begin{align}
|S| < S_{\rm cr} \equiv \ro{\frac{m^2}{\lambda}}.
\end{align}
In the absence of the thermal corrections, therefore, $S$ must be outside this region throughout the evolution
of the Universe since otherwise it would end up with the SUSY minimum. 

Gaugino masses are calculated to be
\begin{align}
m_{\lambda} = \frac{g^2}{(4 \pi)^2} \frac{F_S}{\langle S \rangle} 
= \frac{g^2}{(4 \pi)^2} \frac{2 \ro{3} m^2 M_{\rm pl}}{\Lambda^2} .
\end{align}
This simple model has three parameters $m, \Lambda$ and $\lambda$.
In following analysis we fix the ratio $m / \Lambda$
so that the gaugino mass should be $\mathcal{O} (100) {\rm GeV}$.
In particular, we fix the Bino mass $m_{\tilde B} \simeq 300 {\rm GeV}$.\footnote{
The SM-like Higgs boson with mass $\sim 125$\,GeV was recently found by the ATLAS~\cite{:2012gk} and 
CMS~\cite{:2012gu} experiments. For the Bino mass adopted in the text, the light Higgs boson mass
may not be explained, unless the Higgs sector is extended or extra vector-like 
matter~\cite{Moroi:1991mg,Endo:2011xq} is introduced.
Such extensions will not alter the following analysis significantly as we mainly focus on the SUSY breaking
and mediation sectors.
}
Once the gaugino mass is fixed, we can relate the gravitino mass to $\Lambda$ as
\bea
m_{3/2} \;\simeq\; 0.6{\rm\,GeV} \lrf{m_{\tilde B}}{300{\rm\,GeV}} \lrfp{\Lambda}{10^{16}{\rm\,GeV}}{2}.
\label{m32}
\eea

\section{Pseudo-modulus dynamics}
\label{sec:3}

\subsection{Vacuum selection}
As the model breaks SUSY at a meta-stable state, we have to
check whether the  SUSY  breaking minimum is actually selected
in the cosmological evolution.  The vacuum selection in meta-stable
SUSY breaking models has been discussed in several literature for the
ISS-type models~\cite{Abel:2006cr,Craig:2006kx,Fischler:2006xh,Abel:2006my,
Anguelova:2007at,Papineau:2008xf,Auzzi:2010wm}
and the generalized O'Raifeartaigh
models~\cite{Katz:2009gh,Moreno:2009nk,Ferrantelli:2009zv,Dalianis:2010yk,Dalianis:2010pq}.
In particular, it was shown in Ref.~\cite{Dalianis:2010pq} that the
pseudo-modulus successfully moves from near the origin to the SUSY breaking
minimum in the model of Eqs.~(\ref{kaler}) and (\ref{super}), by virtue
of the finite temperature potentials.  Here we follow their discussion
and clarify the parameter region where $S$ successfully reaches SUSY
breaking vacuum.

We assume that the minimal supersymmetric standard model (MSSM)
superfields and the messenger superfields are in thermal equilibrium in the early Universe.
 Although $S$ superfield is not in thermal plasma, its scalar potential receives   thermal corrections
due to interactions with the
messenger fields. The minimum of the thermal potential is located at the origin,
$\langle S \rangle = \langle f \rangle = \langle \bar f \rangle = 0$,
rather than the SUSY breaking vacuum in Eq.~(\ref{SUSYbmin}).
Thus, for a sufficiently high temperature, the potential minimum is close to the origin, and 
its real component will gradually approaches the SUSY breaking one as the temperature decreases,
whereas the imaginary component remains stabilized at the origin. 

During inflation, on the other hand, there is no thermal plasma. Since the inflaton potential
largely breaks SUSY, the scalar potential of $S$ is modified through Planck-suppressed couplings.
In particular, the $S$ field generically acquires a so-called Hubble-induced mass, and as long as the U(1)$_R$
remains a good symmetry during inflation, the origin of $S$ is close to the extremum of the potential. If the Hubble-induced
mass is positive,  therefore, $S$ is  stabilized near the origin during inflation, and
it likely remains there even after the inflation, which realizes the initial condition mentioned above.

Note that this initial condition is different from that
of Ref.~\cite{Hamaguchi:2009hy}, where the $S$ field is assumed to be
away from the origin in order not to fall into the
SUSY preserving vacuum. In particular, it was  pointed out that the
imaginary part of $S$ needs to have a large initial value to avoid falling into the wrong vacuum, and the
oscillation of the imaginary part is eventually the most important because of its relatively long life time.
In the present case, since the initial position of $S$ is close to the origin after inflation, the imaginary
component of $S$ stays at the origin throughout,
which allows us to concentrate only on the real component of $S$.

The pseudo-modulus $S$ acquires a finite temperature potential  from 
interactions with the messenger fields.  The relevant terms along $f=\bar{f} = 0$ 
direction are calculated up to $\mathcal{O} (S^3)$:
\begin{align}
V_S =
-\frac{2}{\ro 3} \frac{m^4}{M_{\rm pl}} ( S + S^{\dagger} ) + 4 \frac{m^4}{\Lambda^2} |S|^2
+ \frac{5}{4} \lambda^2 T^2 |S|^2 - \frac{5}{3 \pi}  \lambda^3 T |S|^3.
\label{potentialS}
\end{align}
See Appendix~\ref{app:thermal} for the derivation.  The potential
minimum at temperature $T$ is therefore given by
\begin{align}
S_{\rm min} (T) \simeq \frac{\ro{3}}{6} \frac{\Lambda^2}{M_{\rm pl}} 
\frac{m_S^2}{m_S^2 + \frac{5}{4} \lambda^2 T^2} ,
\label{SminT}
\end{align}
where
\begin{align}
m_S \;=\; \frac{ 2 m^2}{\Lambda}
\end{align}
is the tree-level mass of $S$ at zero-temperature.  We can see from this formula
that the potential minimum is near the origin for a high temperature and
moves toward the SUSY breaking one as the
temperature decreases. 

In the absence of the thermal potential, $S$ would fall into the SUSY preserving vacuum if
 $|S| < S_{\rm cr}$, since the  messenger fields become tachyonic.  For a sufficiently high
 temperature, however,  $S$ does not fall into the SUSY preserving vacuum 
 because the thermal effects lift the messenger direction.  The messenger fields get thermal potentials through interactions
with the standard model gauge bosons,
\begin{gather}
V_{\ell} =
-\lambda m^2 ( \ell \bar{\ell} + {\rm h.c.}) + \lambda^2 |\ell|^2 |\bar{\ell} |^2 + \lambda^2 |S|^2 ( |\ell|^2 + |\bar{\ell} |^2 )
+ \frac{T^2}{16} (3 g^2 + g^{\prime 2} ) (|\ell |^2 + | \bar{\ell} |^2 ), 
\label{thermall} \\
V_{q} =
-\lambda m^2 ( q \bar{q} + {\rm h.c.}) + \lambda^2 |q|^2 |\bar{q} |^2 + \lambda^2 |S|^2 ( |q|^2 + |\bar{q} |^2 )
+ \frac{T^2}{16} (8 g_s^2 + g^{\prime 2} ) (|q |^2 + | \bar{q} |^2 ),
\label{thermalq}
\end{gather}
where $\ell$ and $q$ denote the scalar components of the messenger field
$f$.
One can see from the potential that the messenger direction becomes unstable at $S \approx 0$,
when the temperature becomes lower than $T_{\rm cr}$:
\begin{align}
T_{\rm cr} = 
4 m \ro{\frac{\lambda}{3 g^2 + g^{\prime 2}}} .
\end{align}
As $S$ develops a VEV, the critical temperature goes down.

We also define a temperature $T_S$ at which $S$ field exits the region
where the messenger direction is unstable at zero temperature, i.e.,
\begin{gather}
S_{\rm min} (T_S) = S_{\rm cr},
\end{gather}
which gives
\begin{gather}
T_S^2 \simeq 
\frac{8 \ro{3}}{15}
\frac{m^3}{\lambda^{3/2} M_{\rm pl}} .
\end{gather}

\begin{figure}[t]
\begin{center}
\includegraphics[width=8cm,bb=0 0 595 582]{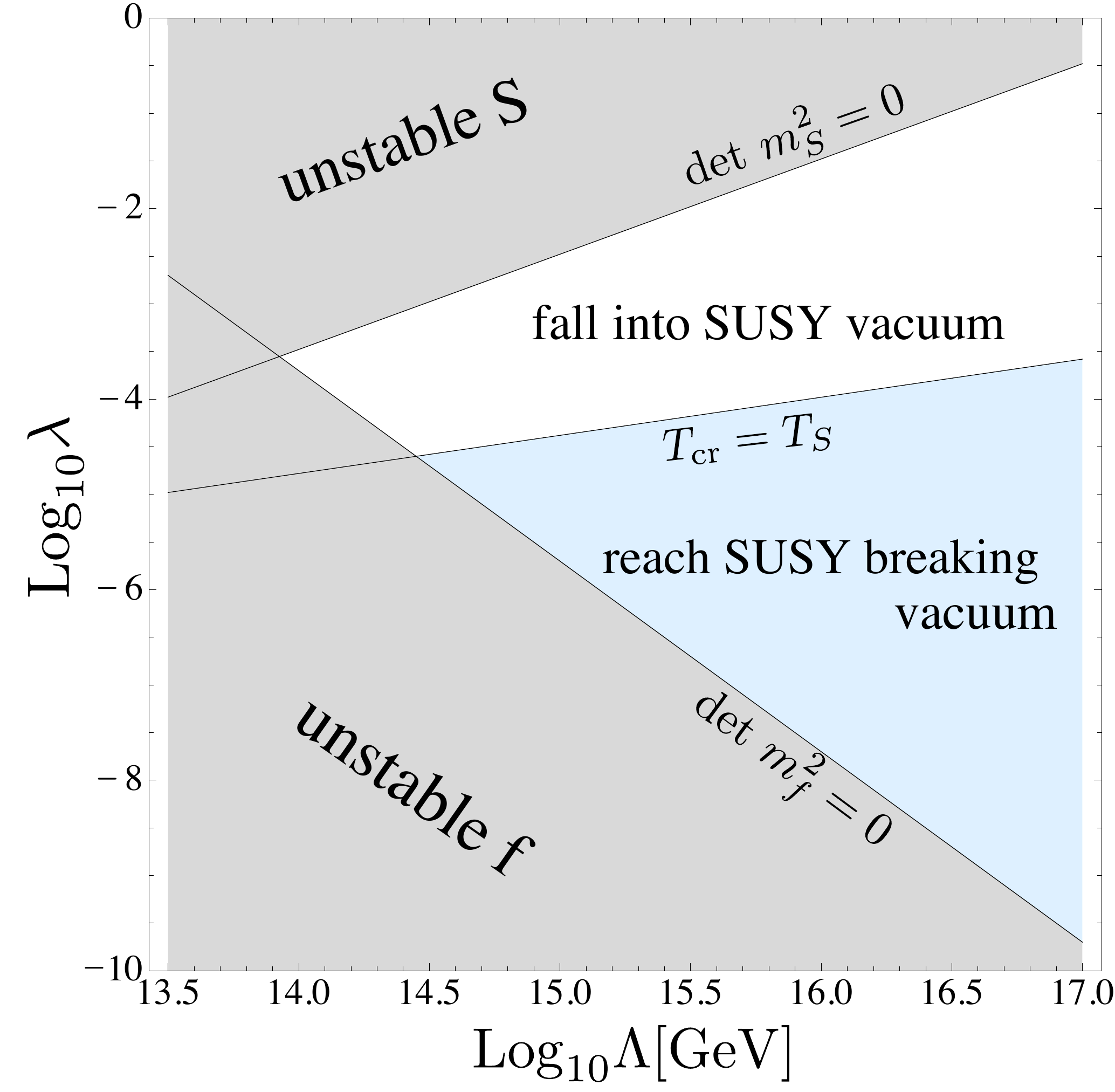} 
\caption{
SUSY breaking vacuum $\langle S \rangle$
appears in the white and blue regions.
Starting from the origin at high temperature, $S$ field successfully reaches the SUSY breaking vacuum
in the blue region.
The model parameter $m$ is taken to be $m = 1.7 \times 10^{-7} \Lambda$ 
so that the Bino mass is fixed to be $300 {\rm GeV}$.
}
\label{allowed}
\end{center}
\end{figure}

In order for the $S$ field to reach the SUSY breaking vacuum without
falling into the SUSY preserving one, the condition
\begin{align}
T_S > T_{\rm cr}
\end{align}
has to be satisfied\footnote{Precisely speaking, this is a sufficient condition, but we have confirmed that
this is consistent with the numerical results.},
 i.e., the $S$ field should leave the dangerous region
(the vicinity of $S=0$) before the messenger direction becomes
tachyonic.  This condition is converted to a constraint on the model
parameters,
\begin{align}
\lambda < 
\left[
\frac{8 \ro{3}}{15} \frac{3g^2 + g^{\prime 2}}{16} \frac{m}{M_{\rm pl}} 
\right]^{2/5} .
\end{align}
We show the allowed region in Fig.~\ref{allowed}.
From now on, we focus the discussion on the blue region where the pseudo-modulus successfully reaches the SUSY breaking
vacuum.

\subsection{Coherent oscillations}
Let us examine the evolution of $S$ direction more closely.  We will
find that there are various  qualitatively different possibilities for
the $S$ motion after the inflation.

During  inflation, we assume that $S$ was stabilized near the origin by the
positive Hubble-induced mass term. After inflation, the $S$ follows the 
time-dependent minimum (\ref{SminT}). When the thermal mass  becomes 
comparable to the tree-level mass, the minimum $S_{\rm min} (T)$ quickly moves to the SUSY
breaking minimum. This transition takes place at $T \simeq T_0$, given by
\begin{align}
T_0 \equiv \frac{4}{\ro{5}} \frac{m^2}{\Lambda \lambda}.
\end{align}
Whether or not the $S$ field catches up the motion of the minimum
depends on the competition between the effective mass of $S$ and
the friction caused by the expansion of the Universe. Here it is assumed that the messenger fields
remain thermalized at $T=T_0$.  Later we study the case where the
messenger fields decouple from the thermal plasma before the temperature
of the Universe goes down to $T_0$.

The dynamics of $S$ field is governed by the equation of motion
\begin{align}
\ddot{S} + 3H \dot{S} + \frac{\partial}{\partial S^{\dagger}} V = 0.
\end{align}
The effective mass of $S$ field  is approximately given by the sum of the tree-level mass and the
thermal mass,
\begin{gather}
m_S^2 (T) \equiv m_S^2 + \frac{5}{4} \lambda^2 T^2,
\label{mST}
\end{gather}
where we have neglected the contribution from the radiative correction to the K\"ahler potential (\ref{kaler1}),
since it does not affect results. In the numerical calculations, the effect of the radiative correction is
properly taken into account.

First let us consider the case that the Hubble parameter is  larger
than the effective mass at $T=T_0$, i.e., $H (T_0) > m_S (T_0)$.  In
this case, even if the potential minimum moves to the zero temperature
value at $T=T_0$, $S$ is still trapped near the origin because of the
large friction.  At a later time when the Hubble parameter becomes
comparable to the effective mass, $H \sim m_S$, $S$ leaves the vicinity of the origin
and starts  oscillations about the minimum.  We define the
temperature $T_{\rm osc}$ as
\begin{align}
H (T_{\rm osc} ) = m_S (T_{\rm osc} ),
\end{align}
where the temperature dependence of the Hubble parameter is given by
$H(T) \sim \frac{T^2}{M_{\rm pl}}$ and
$H(T) \sim \frac{T^4}{M_{\rm pl} T_R^2}$
in the radiation and inflaton-matter dominated eras, respectively.
The condition $H (T_0) > m_S (T_0)$ is equivalent to $T_0 > T_{\rm
osc}$.

\begin{figure}[h!t]
\begin{center}
\includegraphics[width=8cm,bb=0 0 408 283]{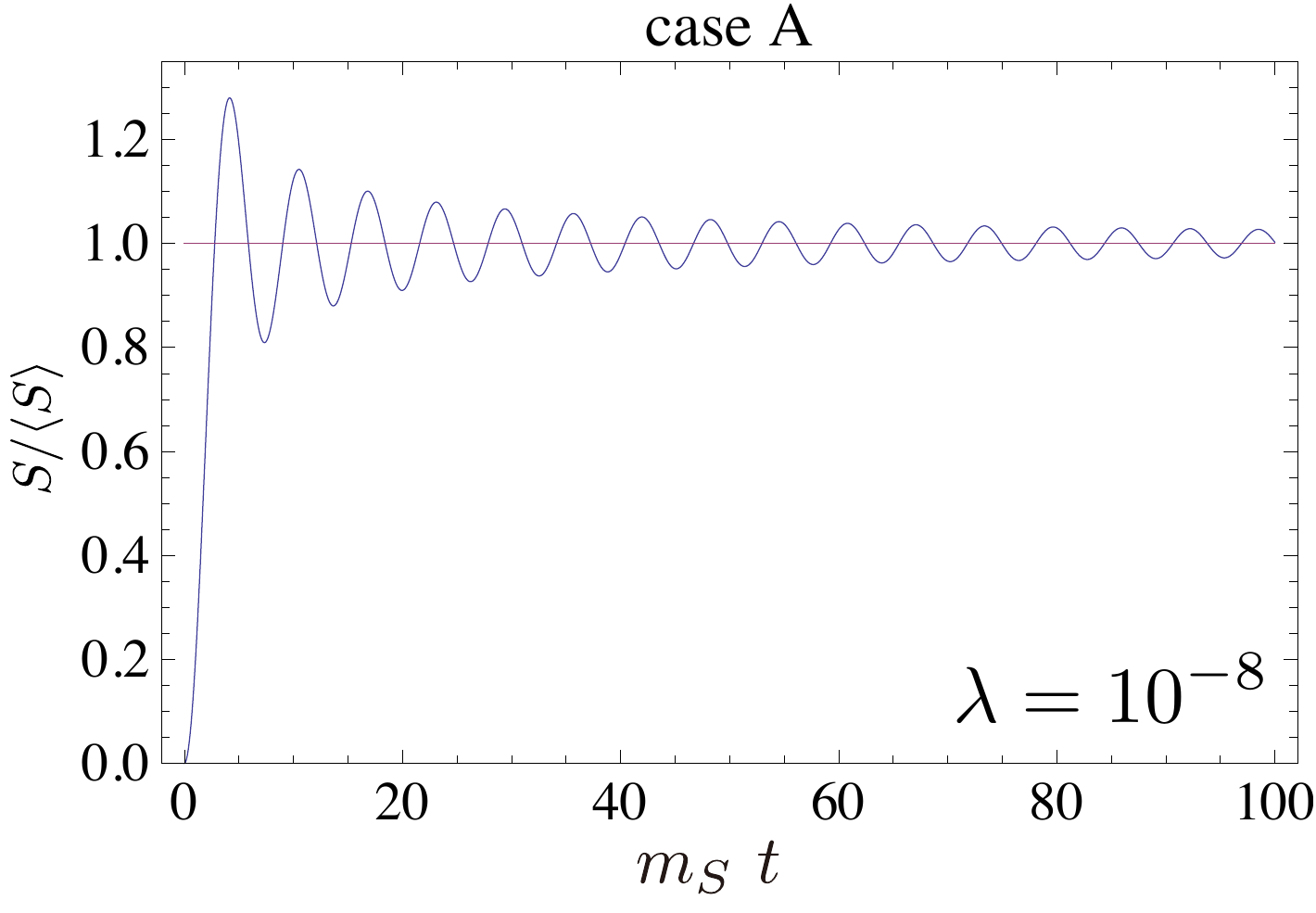}
\includegraphics[width=8cm,bb=0 0 408 283]{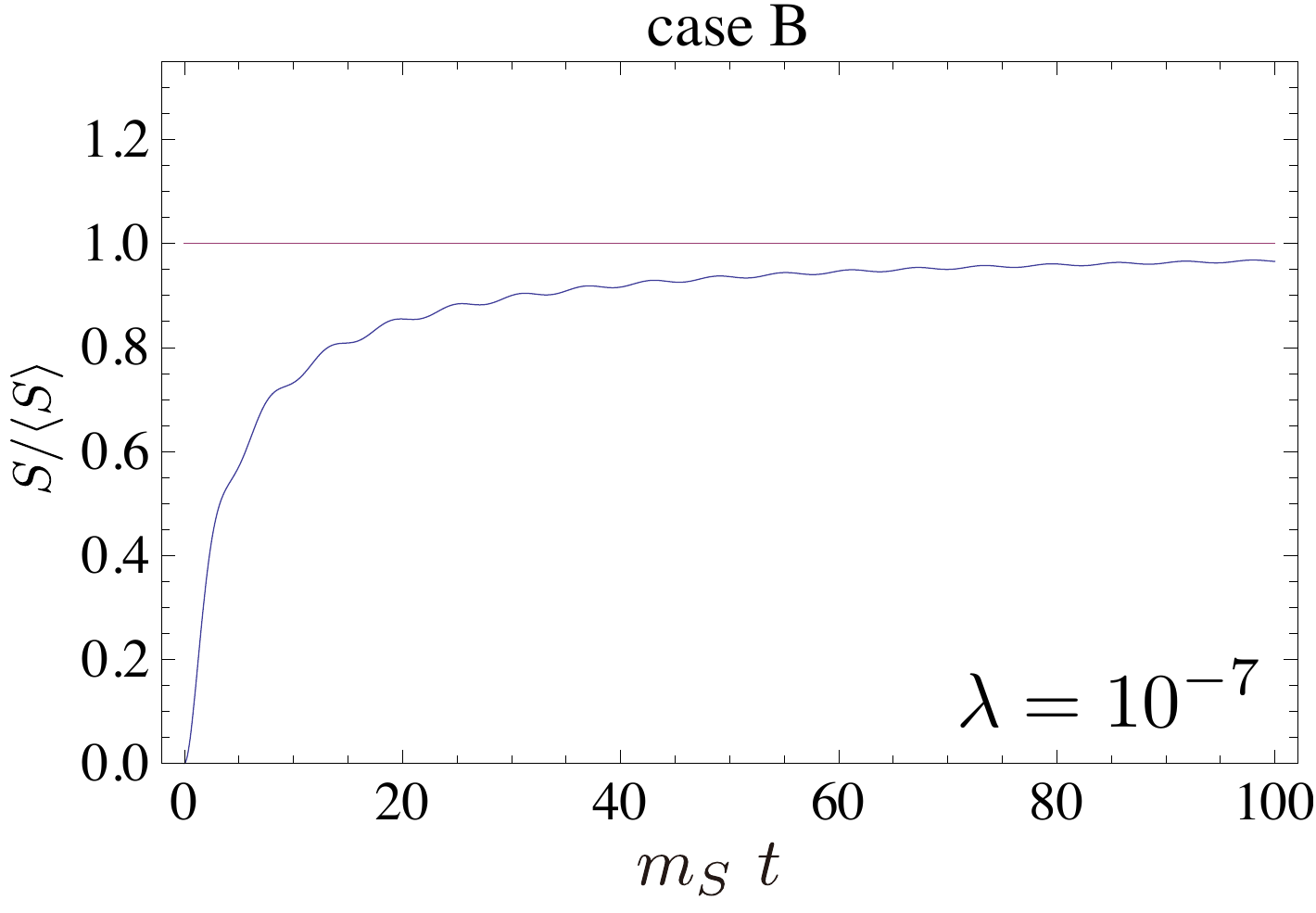}
\end{center}
\caption{
If $m_S (T_0) < H (T_0)$, 
$S$ starts coherent oscillations around $\langle S \rangle$
when $H \sim m_S$ (case A). The thermal mass increases as $\lambda$,
and the adiabatic suppression takes place (case B). The bottom panel
show the case of $m_S (T_0) > H(T_0)$,  where one can see that
$S$ follows the minimum $S_{\rm min} (T)$ with suppressed oscillations.
We set $\Lambda = 10^{16.5}$\,GeV.
}
\label{linde}
\end{figure}

On the other hand, if the Hubble parameter is already smaller than the
effective mass at $T=T_0$, $H (T_0) < m_S (T_0)$, or equivalently $T_0 <
T_{\rm osc}$, the friction from the expansion of the Universe is small.
This is the case if $\lambda$ is larger than the previous case.
Then the $S$ field follows the time-dependent potential minimum and gradually
reaches the SUSY breaking vacuum.  The amplitude of oscillations is
highly suppressed in this case. The suppression was first found in Ref.~\cite{Linde:1996cx},
in which  the oscillation amplitude was shown to be exponentially suppressed
in a limiting case\footnote{The initial condition adopted in Ref.~\cite{Linde:1996cx}
was given at an infinitely large Hubble parameter. There are various additional contributions
in general, which are only power-suppressed~\cite{Nakayama:2011wqa}.
We have numerically confirmed that the pseudo-modulus abundance is power-suppressed in our scenario.
} in the context of the cosmological moduli problem.
This adiabatic suppression mechanism was recently examined more carefully 
in Ref.~\cite{Nakayama:2011wqa}.  
We show in Fig.~\ref{linde} the typical evolution of $S$ in the above two cases.

\begin{figure}[t]
\begin{center}
\includegraphics[width=8cm,bb=0 0 417 283]{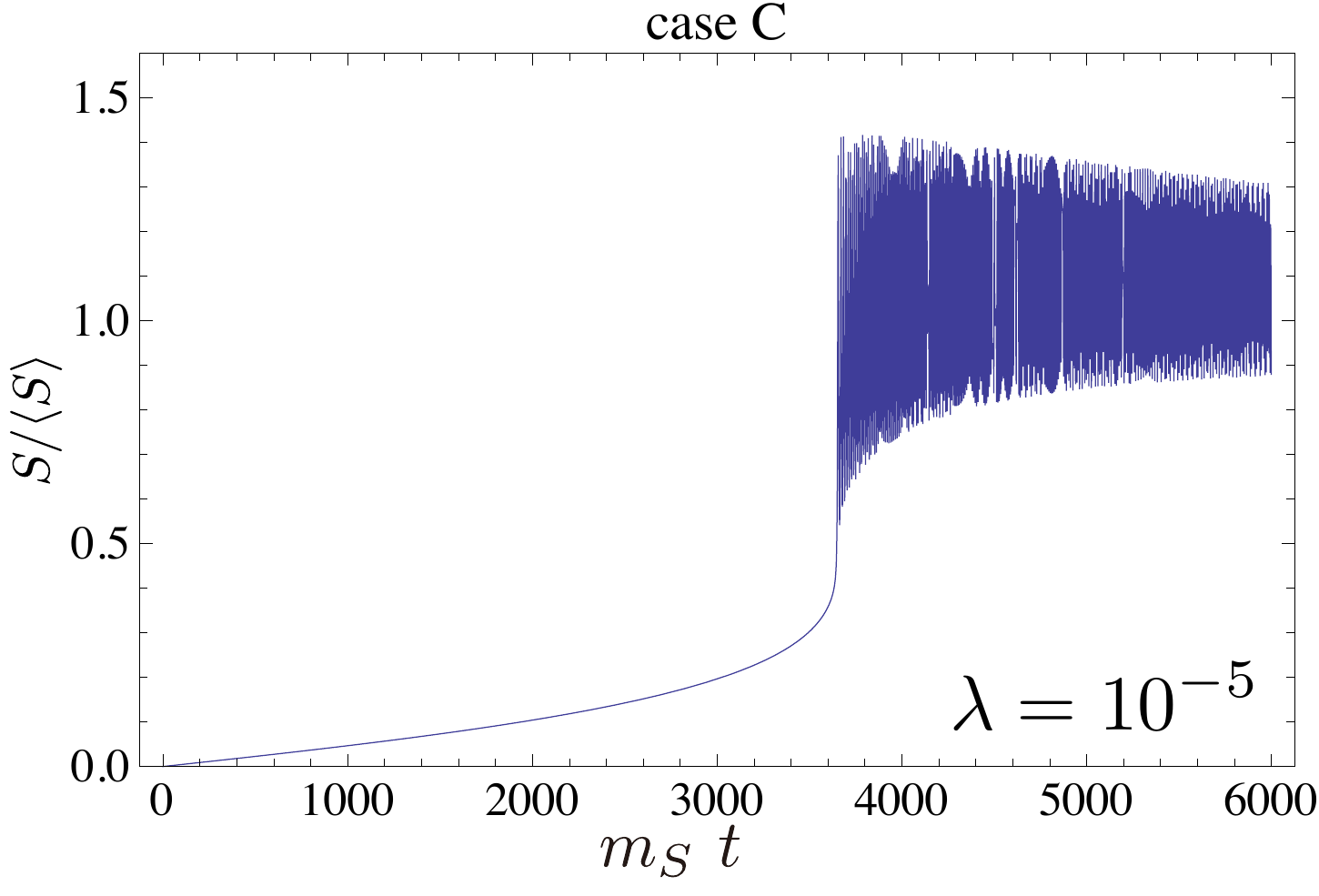}
\caption{
If the messenger fields become non-relativistic before $S$ reaches the
zero temperature SUSY breaking vacuum $\langle S \rangle$,
the thermal mass term quickly disappears and $S$ field begin to oscillate about $\langle S \rangle$.
We set $\Lambda = 10^{16.5}$\,GeV.
}
\label{messenger_decouple}
\end{center}
\end{figure}

There is yet another possibility.
As the temperature of the
Universe decreases and the value of the pseudo-modulus becomes sizable, the
high temperature approximation in Eq.(\ref{potentialS}) breaks down at
a certain point.   The messenger fields become non-relativistic when the temperature of the
Universe becomes comparable to the messenger mass.  Then the finite
temperature potential generated by messenger interactions gets suppressed by
 the Boltzmann  factor $\sim e^{- \lambda S / T}$.  We define
the decoupling temperature $T_{\rm dec}$ as
\begin{align}
T_{\rm dec} \equiv \lambda S(T=T_{\rm dec} ).
\label{Tdec}
\end{align}
If the messenger fields decouple after $S$ field reaches the SUSY breaking vacuum, 
$T_{\rm dec}$ is simply the  messenger mass scale,
\begin{align}
 T_{\rm dec} \;\simeq\; \lambda \langle S \rangle 
 \end{align}
for $T_{\rm dec} < T_0$.
If the decoupling occurs when $S$ is still on the way to the SUSY breaking vacuum,
$T_{\rm dec}$ is calculated to be
\begin{align}
T_{\rm dec} 
\;\simeq\; \left[ \frac{8 \ro{3}}{15} \frac{m^4}{\lambda M_{\rm pl}} \right]^{1/3}
\label{Tdec2}
\end{align}
for $T_{\rm dec} > T_0$.
In this case, at $T=T_{\rm dec}$, the position of the potential minimum instantly moves from $S_{\rm min} (T)$ to the SUSY breaking
vacuum, which triggers coherent oscillations about the minimum (see Fig.~\ref{messenger_decouple}).

In summary, we have defined three temperatures: $T_0$, $T_{\rm osc}$ and
$T_{\rm dec}$.
\begin{itemize}
\item{
The potential minimum quickly moves from the origin to the SUSY
     breaking vacuum at $T=T_0$.
}
\item{The Hubble parameter $H$ becomes comparable to the pseudo-modulus mass $m_S(T)$ at $T=T_{\rm osc}$.}
\item{The messenger fields become non-relativistic and disappear from the thermal plasma at $T=T_{\rm dec}$.}
\end{itemize}

\begin{figure}[t!]
\begin{center}
\includegraphics[width=7cm,bb=0 0 556 577]{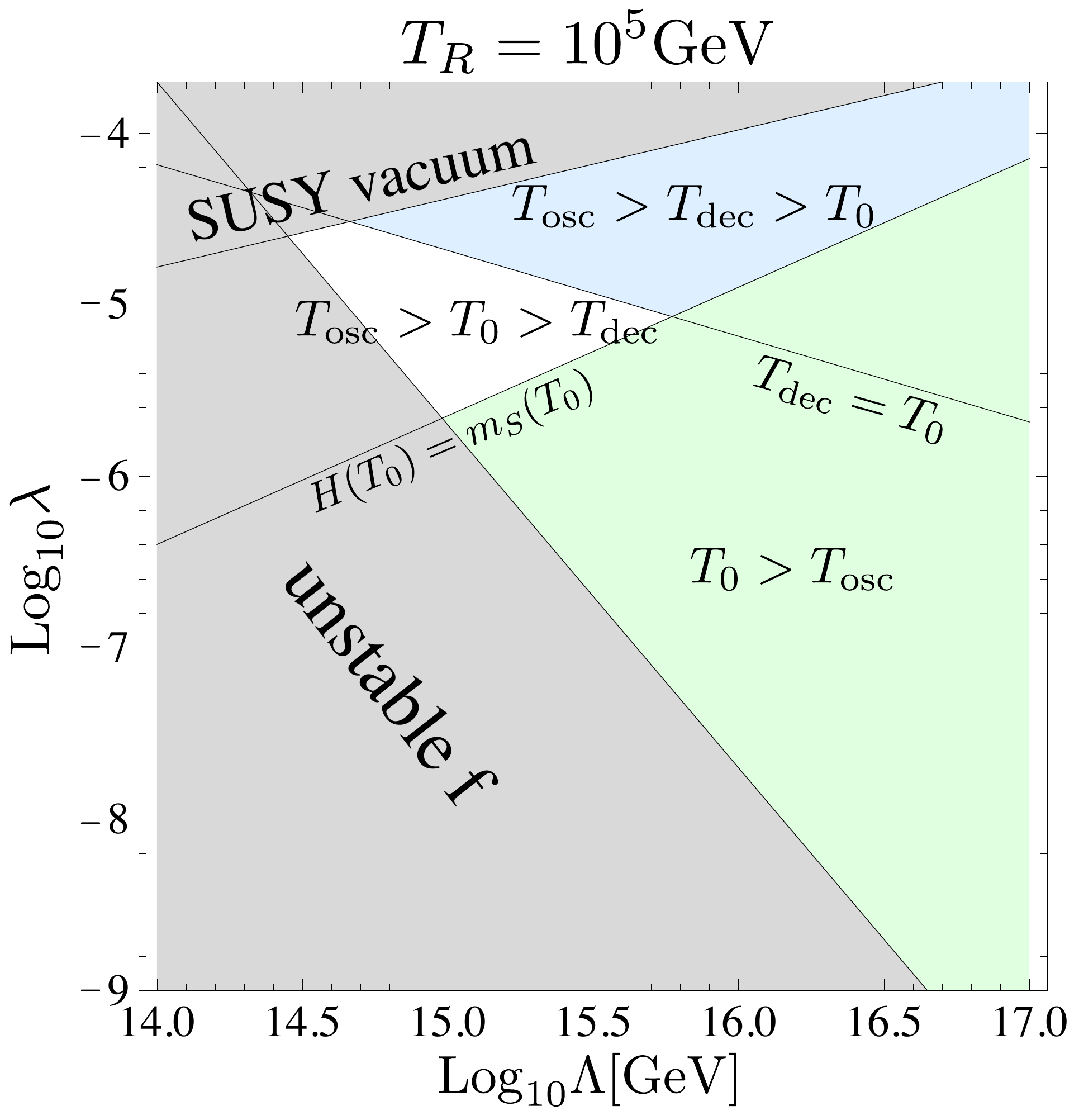} 
\includegraphics[width=7cm,bb=0 0 556 566]{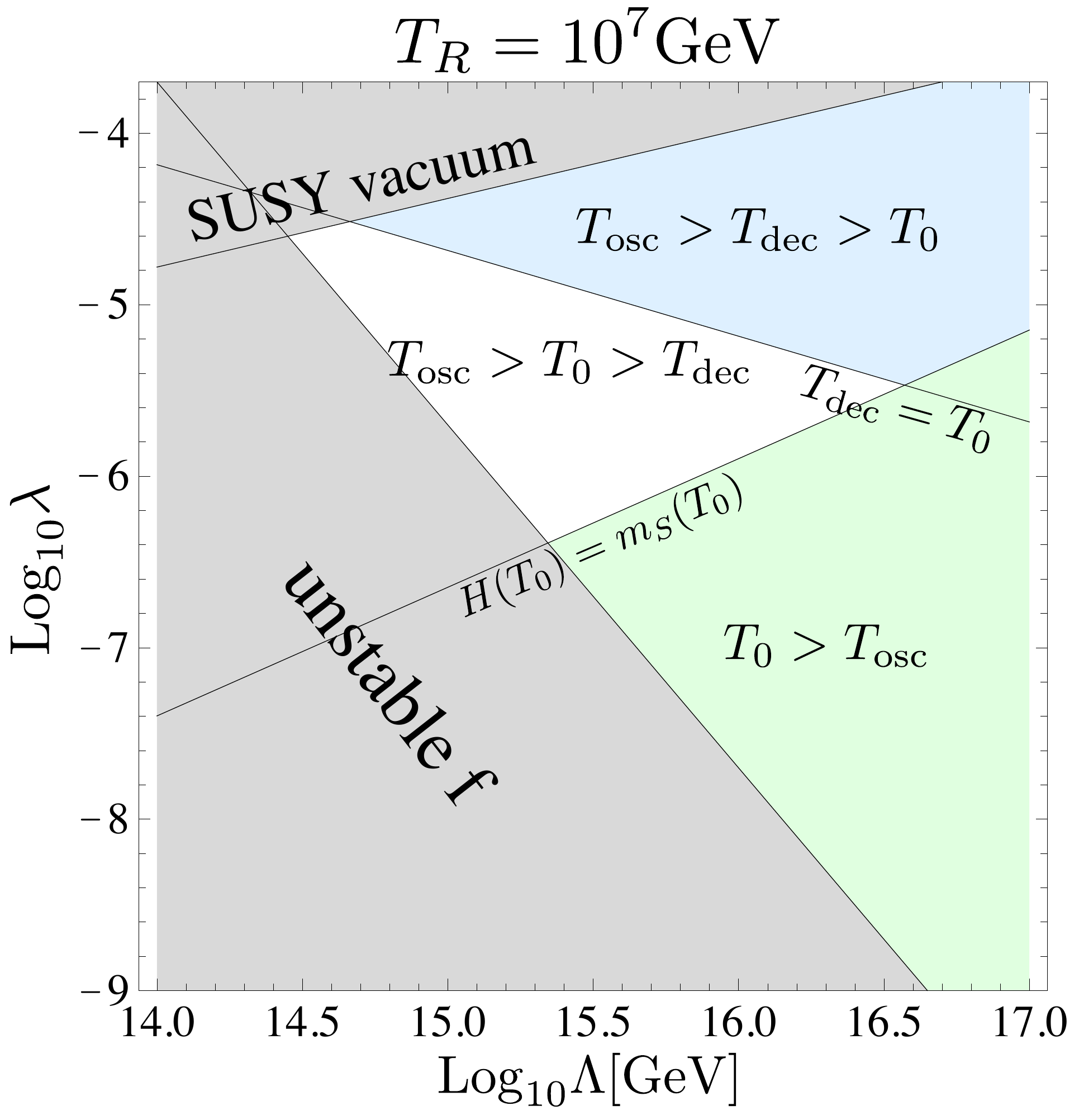} 
\includegraphics[width=7cm,bb=0 0 555 566]{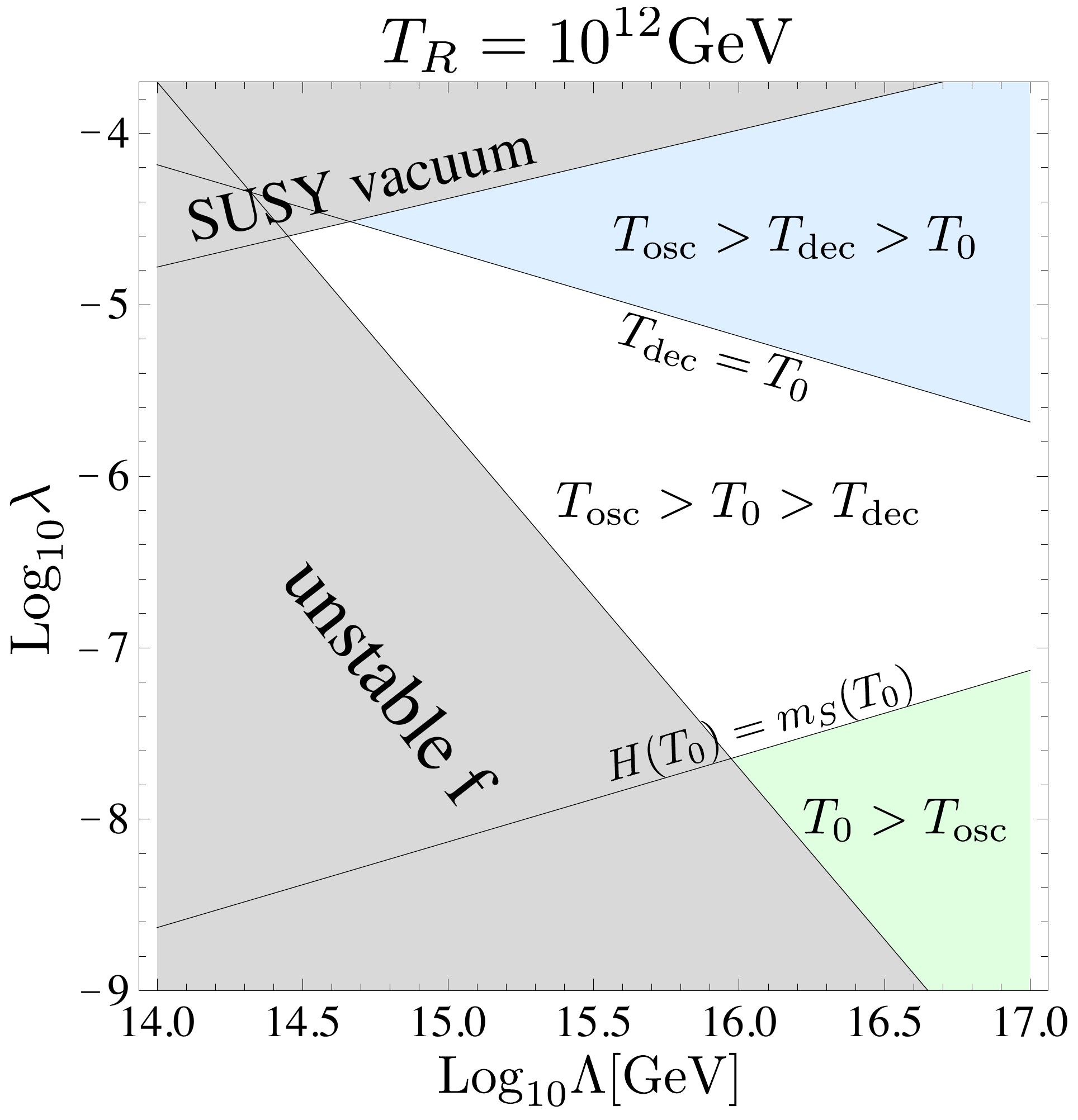} 
\caption{ The evolution of $S$ exhibits
distinctive behavior depending on the model parameters.
In the green region ($T_0 > T_{\rm osc}$) $S$ starts coherent oscillations
when $H \approx m_S(T)$.
The $S$ abundance is suppressed by the adiabatic suppression mechanism in the white region
($T_{\rm osc} > T_0 > T_{\rm dec}$).  In the blue region ($ T_{\rm osc}
> T_{\rm dec} > T_0$) the coherent oscillations are triggered when messenger fields decouple from thermal 
plasma and disappear.
While $T_0$ and $T_{\rm dec}$ are uniquely determined once we fix the model parameters, 
the value of $T_{\rm osc}$ depends on $T_R$.
} \label{allowed2}
\end{center}
\end{figure}

The evolution of $S$, and therefore, its abundance, sensitively depends on the relations among these
temperatures. The following three cases are shown in Fig.~\ref{allowed2} for several reheating temperatures. 
Note that the abundance of $S$ is suppressed in the second case (white region), while it is significant
in the other cases. 
\begin{itemize}
\item{$T_0 > T_{\rm osc}$ (Green region) ; $S$ starts coherent oscillation about $\langle S \rangle$ at $T=T_{\rm osc}$.}
\item{$T_{\rm osc} > T_0 > T_{\rm dec}$ (White region) ; $S$ follows $S_{\rm min} (T)$ and gradually
reaches $\langle S \rangle$ without sizable oscillations.}
\item{$T_{\rm osc} > T_{\rm dec} > T_0$ (Blue region) ; The messenger fields decouple from thermal plasma when $S$
is on the way to $\langle S \rangle$. Coherent oscillations are triggered at $T=T_{\rm dec}$.}
\end{itemize}

\section{Thermal production of the gravitinos}
\label{sec:4}
In the framework of gauge mediation the dominant component of thermally produced gravitino 
is that of the longitudinal mode, the goldstino.
The relic abundance is estimated to be~\cite{Moroi:1993mb,de Gouvea:1997tn,Bolz:1998ek,Bolz:2000fu,Pradler:2006qh,Pradler:2006hh,
Rychkov:2007uq}

\begin{align}
\Omega_{3/2}^{\rm th} \simeq 0.3 \left( \frac{\rm GeV}{m_{3/2}} \right)
\biggl( \frac{m_{\tilde g}}{\rm TeV} \biggr)^2
\left( \frac{T_ R}{10^7 {\rm GeV}} \right).
\label{omega32usual}
\end{align}
where $T_R$ represents the reheating temperature after inflation, and
it is assumed that there is no entropy production after the reheating.
Note however that it is assumed implicitly in deriving this formula that
the messenger mass scale is higher than $T_R$.  
For $T_R$ higher than the messenger scale, one needs to consider diagrams
in which the messenger fields are in the external lines.
As shown in Ref.~\cite{Choi:1999xm}, if $T_R$ is
higher than the messenger mass scale, or equivalently if the messenger
fields once gets thermalized, the goldstino relic abundance is
determined by
the messenger mass scale  $M_{\rm mess} = \lambda \langle S \rangle$ 
rather than the reheating temperature,
\begin{align}
\Omega_{3/2}^{\rm th} \simeq 0.8 \left( \frac{\rm GeV}{m_{3/2}} \right)
\biggl( \frac{m_{\tilde g}}{\rm TeV} \biggr)^2
\left( \frac{M_{\rm mess}}{10^5 {\rm GeV}} \right) \ \ \ \ ( T_R > M_{\rm mess} ).
\label{omega32th}
\end{align}
The replacement of $T_R$ by $M_{\rm mess}$ can be understood by looking
at the temperature dependence of the goldstino reaction rate $\Gamma
(T)$:
\begin{align}
&\Gamma (T)
 \sim \alpha_3 \lambda^2 T,  \ \ \ {\rm for} \ \ \  T \gtrsim M_{\rm mess} \\
&\Gamma (T) 
 \sim \alpha_3 \frac{m_{\tilde g}^2}{m_{3/2}^2 M_{\rm pl}^2} T^3 
 \ \ \ {\rm for } \ \ \ T \lesssim M_{\rm mess}.
\end{align}
As $\Gamma (T) $ has lower power dependence on $T$ than the Hubble
parameter ($ H \propto T^2$) for $T \gtrsim M_{\rm mess}$, contributions
to the goldstino production from MSSM fields decouple for $T \gtrsim
M_{\rm mess}$~\footnote[2]{It has been argued in
Ref.~\cite{Jedamzik:2005ir} that there is a component of the interaction
rate which still grows as $T^3$ at high temperatures, and thus the
estimation of the gravitino abundance is sensitive to $T_R$ in contrast
to the conclusion of Ref.~\cite{Choi:1999xm}. This is based on the
observation that there are contact interactions between the gravitino
and visible sector fields in the supergravity Lagrangian in the unitary
gauge.
However, we anticipate a cancellation of such contributions with loop
diagrams at a high temperature since there is no such contribution in
the goldstino picture.

}.
We should estimate the thermally produced gravitino abundance by Eq.(\ref{omega32th})
since the pseudo-modulus was at the origin and the messenger fields were inevitably thermalized 
in our scenario.

\section{Pseudo-modulus decay and the gravitino abundance}
\label{sec:5}

In this section we examine the decays of the pseudo-modulus.
So far  we have assumed that there is no additional entropy production after reheating.
If the oscillation energy of the
pseudo-modulus  dominates the Universe, however, a large amount of entropy is produced by
the decay, and the pre-existing gravitino is diluted.  Also the
gravitinos are produced by the decay of the pseudo-modulus.

\subsection{$S$-domination}

The energy density of $S$ decreases more slowly than radiation, and so, if its lifetime
is sufficiently long or the reheating temperature is high, $S$ may come to dominate
the Universe and produce a large amount of entropy by the decay.
Let us define the entropy dilution factor $\Delta$ as
\bea
\frac{1}{\Delta} \equiv \frac{s_{\rm inf}}{s_S+s_{\rm inf}} & \simeq & {\rm Min}\left[1,
\frac{s_{\rm inf}}{s_S} \right],
\eea
where $s_{\rm inf}$ and $s_S$ represent the entropy density produced from the inflaton and the
pseudo-modulus, respectively. If $\Delta > 1$, the pre-existing gravitino abundance is 
diluted by a factor of $1/\Delta$. In this case, we can express the dilution factor in terms of 
the decay temperature of $S$, $T_{\rm d}$,  and its abundance as
\bea
\Delta \simeq \frac{s_S}{s_{\rm inf}} =
\frac{4} {3T_{\rm d}} \cdot \left.
\frac{\rho_S}{s_{\rm inf}}
\right|_{S~decay}.
\eea
The pseudo-modulus abundance, $(\rho_S / s_{\rm inf})$, can be estimated by following its evolution 
as discussed in Sec.~\ref{sec:3}. 
The decay temperature $T_{\rm d}$ is defined as
\begin{align}
T_{\rm d} \;\equiv\; \left( \frac{90}{\pi^2 g_{\ast}} \right)^{1/4} \ro{M_{\rm pl} \Gamma_S},
\end{align}
where $\Gamma_S$ is the decay width of $S$.

The decay width of $S$ can be explicitly calculated. 
The $S$ field couples to the MSSM particles through loop diagrams of the
messenger fields. The partial decay width of the two gravitino mode is
suppressed compared to the MSSM particles.
The interaction Lagrangians between $S$ and MSSM fields and calculations
of decay widths can be found in Refs.~\cite{Ibe:2006rc,Hamaguchi:2009hy}.
The main decay mode is $S \to b \bar{b}$ and $S \to hh+WW+ZZ$ for $2 m_b
< m_S < 2 m_W$ and $2 m_W < m_S \lesssim 1 {\rm TeV}$ cases,
respectively.  The decay temperature is calculated to be $\mathcal{O}
(0.1-10) {\rm GeV}$ in the parameter region of our interest. (See
Refs.~\cite{Ibe:2006rc,Hamaguchi:2009hy} for formulae and parameter
dependences.)

\begin{figure}[t!]
\begin{center}
\includegraphics[width=7cm,bb=0 0 555 577]{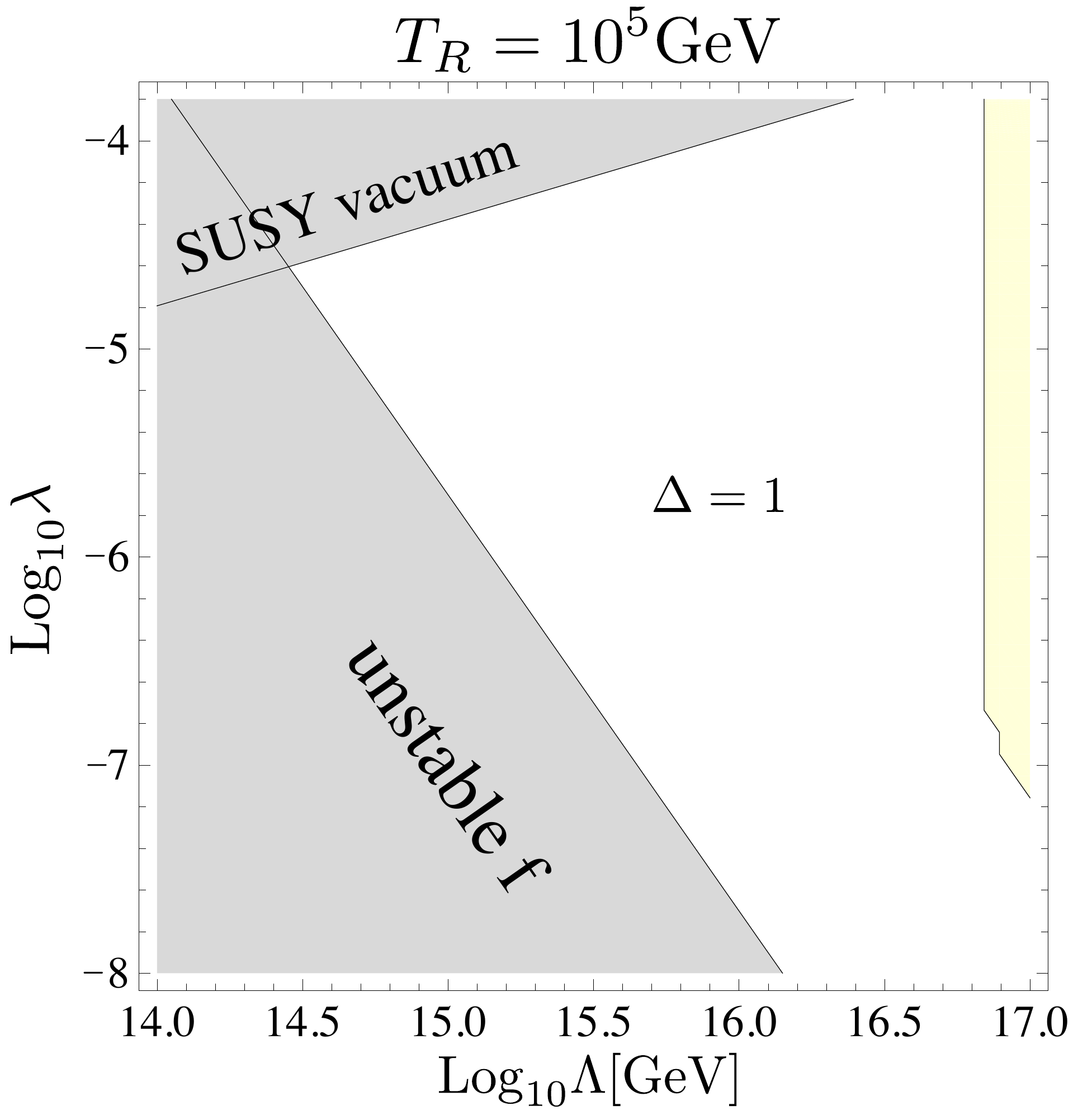} 
\includegraphics[width=7cm,bb=0 0 560 576]{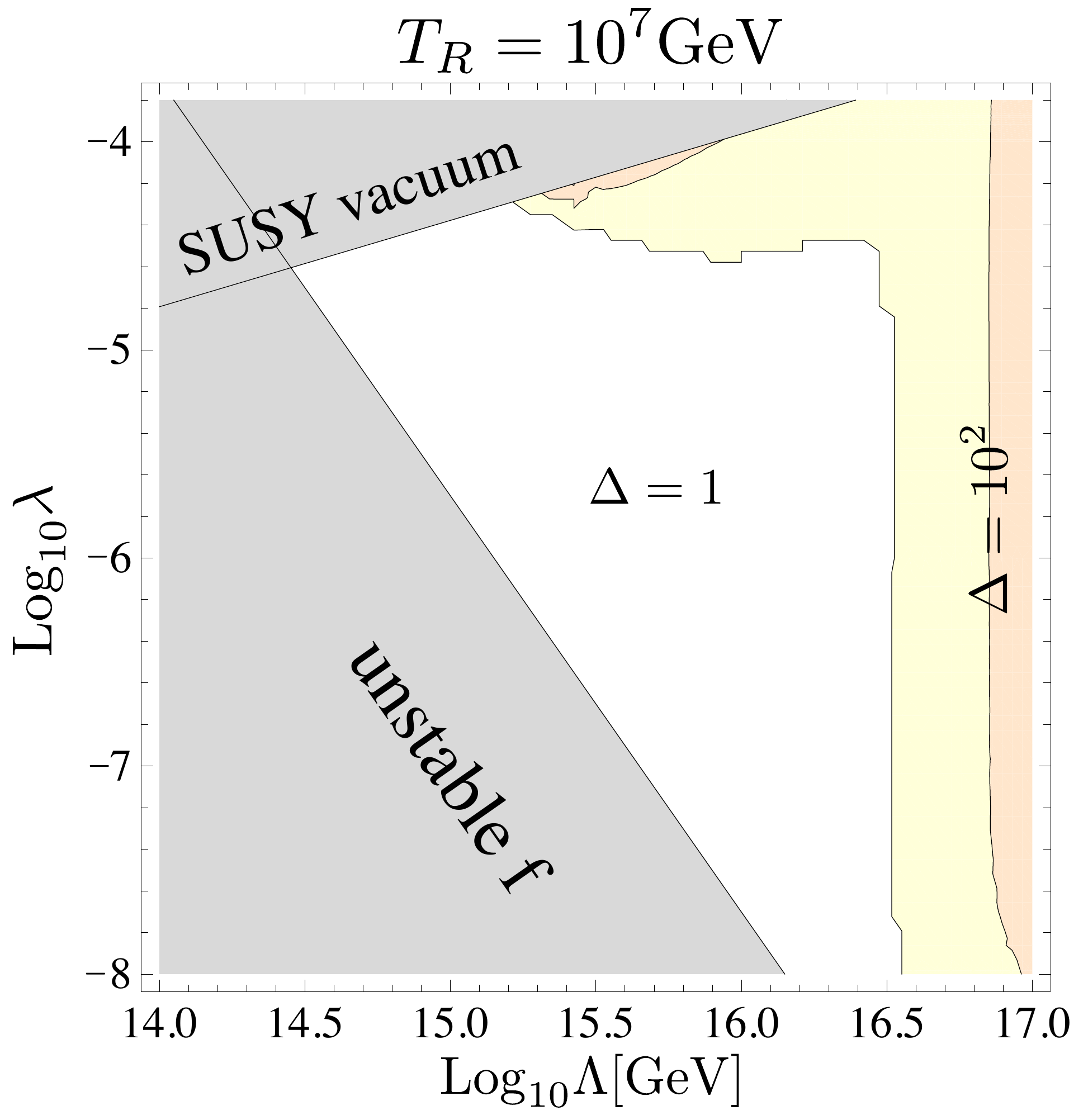} 
\includegraphics[width=7cm,bb=0 0 550 577]{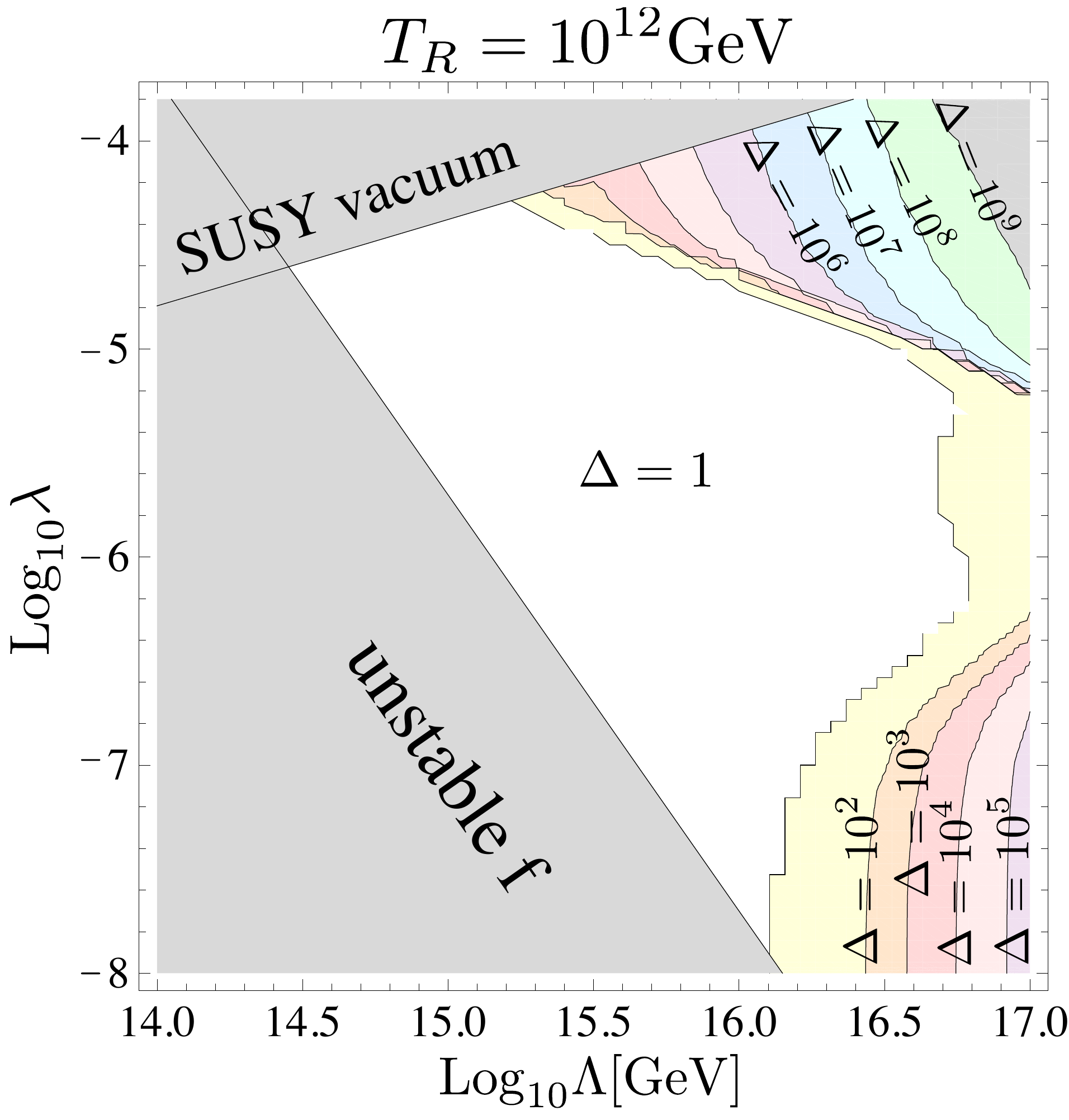} 
\end{center}
\caption{
The contour plot of the dilution factor $\Delta$, for $T_R = 10^5$, $10^7$, and $10^{12}$\,GeV.
The colored regions correspond to the regions where  the pseudo-modulus 
dominates the Universe before the decay.
In the case of $T_R = 10^{12} {\rm GeV}$, the oscillations of $S$ commence after the reheating, 
and the $S$ tends to dominate the  Universe, producing larger entropy compared to the 
other two cases with lower $T_R$.
}
\label{dilution}
\end{figure}

We evaluate the dilution factor numerically. We solve the equation of
motion of the pseudo-modulus which provides the pseudo-modulus abundance,
$\rho_S/s_{\rm inf}$. For the potential at finite temperatures, we
use the one in Eq.~(\ref{VTfull}).
The decay temperature is calculated by using the formulae in
Refs.~\cite{Ibe:2006rc,Hamaguchi:2009hy}.
The results are shown in Fig.~\ref{dilution}. 
The colored regions ($\Delta > 1$) correspond to the regions where the
oscillation energy dominates the energy density of the Universe.
One can see that the $S$ abundance is indeed suppressed in the
middle of each panel, because of the adiabatic suppression mechanism.
The similarity with Fig.~\ref{allowed2} is clear especially in the case of $T_R = 10^{12}$\,GeV.

\subsection{Non-thermally produced gravitino}
Gravitinos are produced non-thermally by the rare decay $S \to
\psi_{3/2} \psi_{3/2}$.
The branching ratio of the gravitino mode, $B_{3/2}$, is calculated in
Refs.~\cite{Ibe:2006rc,Hamaguchi:2009hy}, and it is typically of ${\mathcal
O}(10^{-10}- 10^{-6})$ in the parameter regions of our interest.

If $S$ dominates the energy density of the Universe, 
non-thermal gravitino abundance is calculated as
\begin{align}
\Omega_{3/2}^{\rm NT}
=
\frac{3}{4} m_{3/2} \frac{T_{\rm d}}{m_S} B_{3/2} \times 2 
/
(\rho_c / s)_0 ,
\label{omegant}
\end{align}
where $(\rho_c / s)_0 \simeq 3.6 \times 10^{-9}\,h^{-2} {\rm GeV}$.
We will see later that there are parameter regions where the DM relic
density is explained by the non-thermally produced gravitinos.

\subsection{Total gravitino abundance}
The relic density of the gravitinos is given by the sum of the thermally and non-thermally
produced gravitinos. If the $S$ dominates the Universe before the decay, we need to
take account of the entropy factor, which dilutes the pre-existing gravitinos produced at the 
reheating;
\begin{align}
\Omega_{3/2}^{\rm tot} = \frac{1}{\Delta} \Omega_{3/2}^{\rm th}
+ \Omega_{3/2}^{\rm NT}.
\end{align}
Here the dilution factors are evaluated numerically as in Fig.~\ref{dilution}.
Note that $T_{\rm d}$ is so low that the thermal production
of gravitinos is negligible at the pseudo-modulus decay.

\begin{figure}[h!t]
\begin{center}
\includegraphics[width=7cm,bb=0 0 560 577]{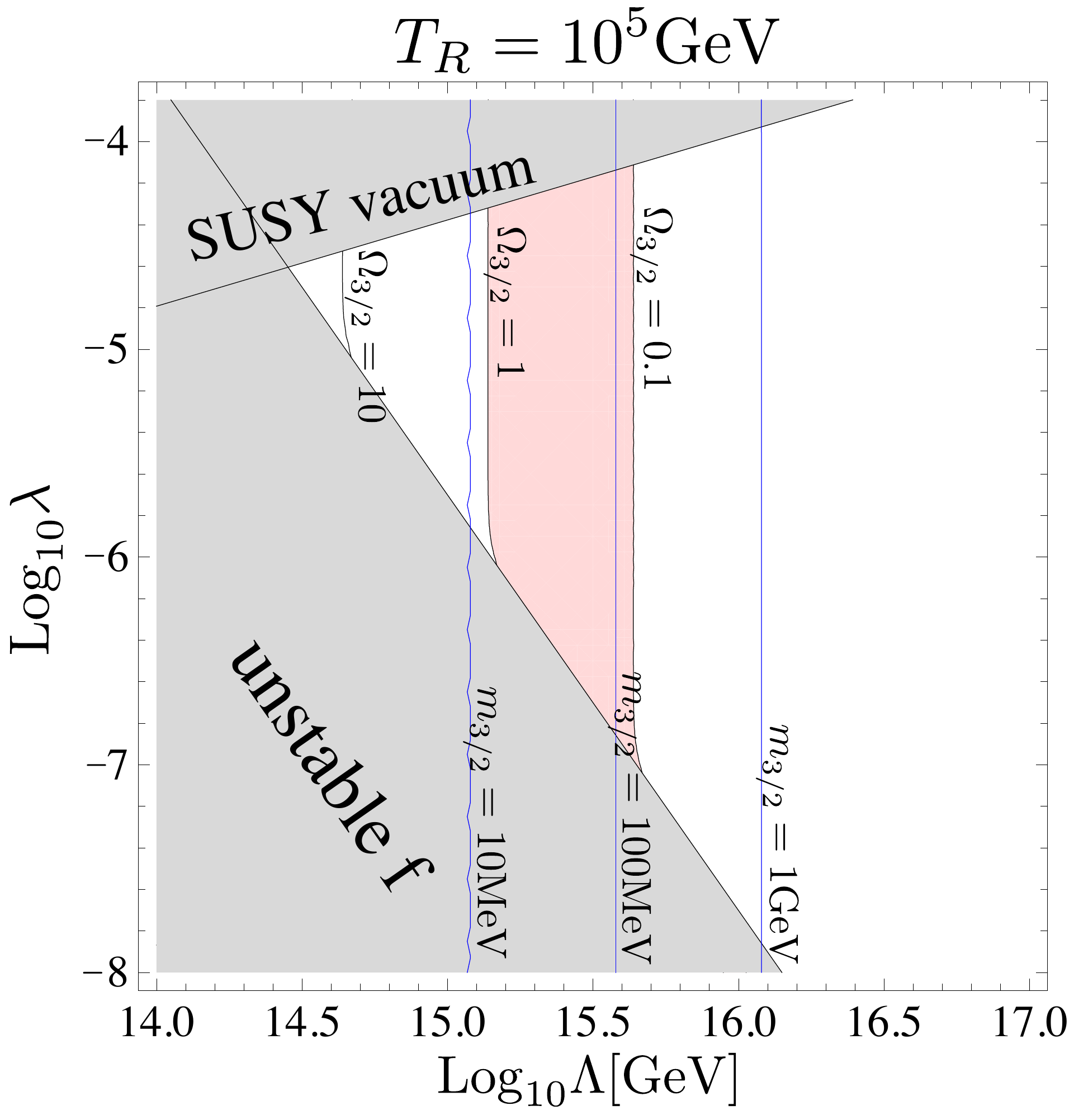} 
\includegraphics[width=7cm,bb=0 0 560 577]{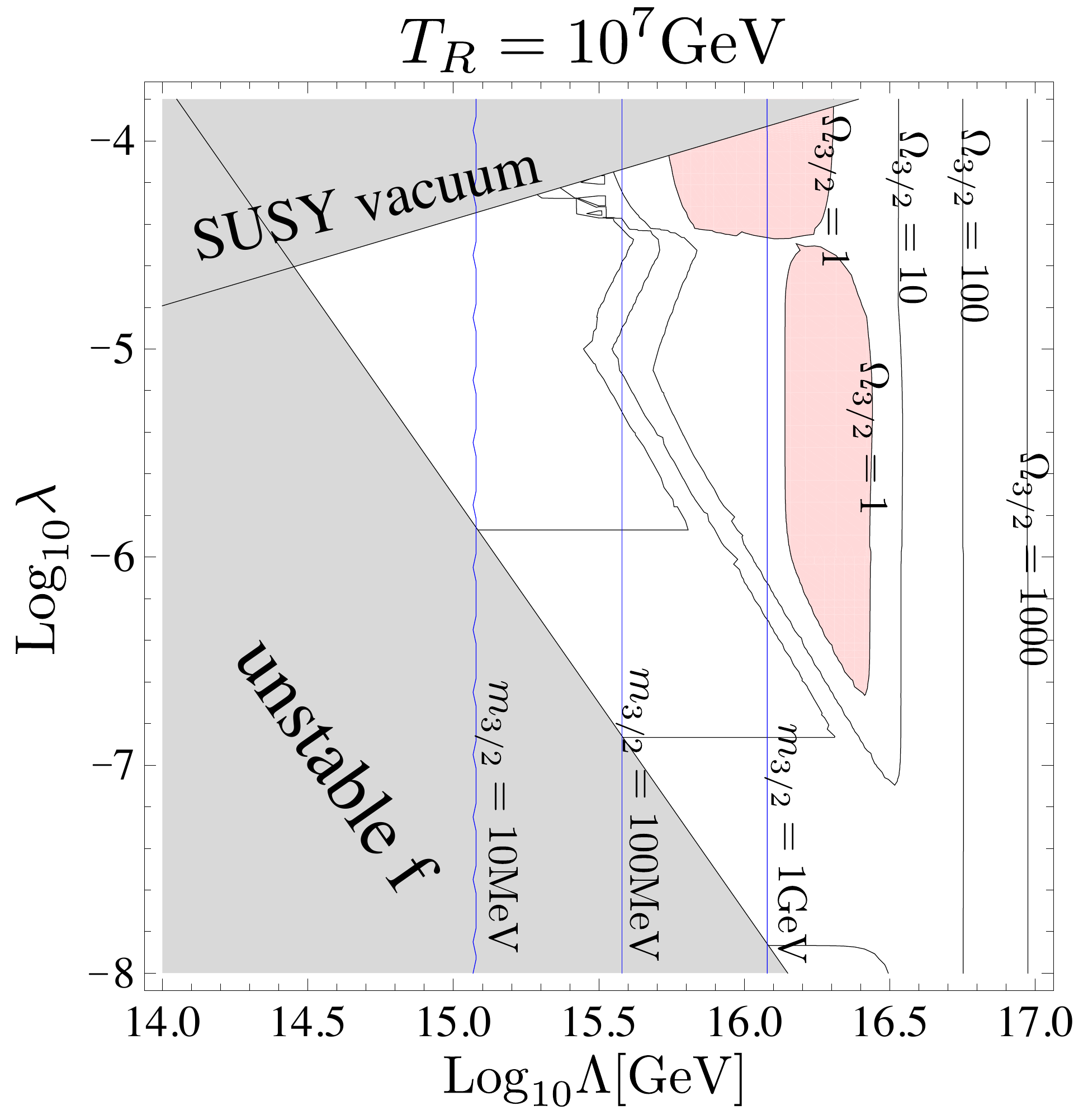} 
\includegraphics[width=7cm,bb=0 0 560 577]{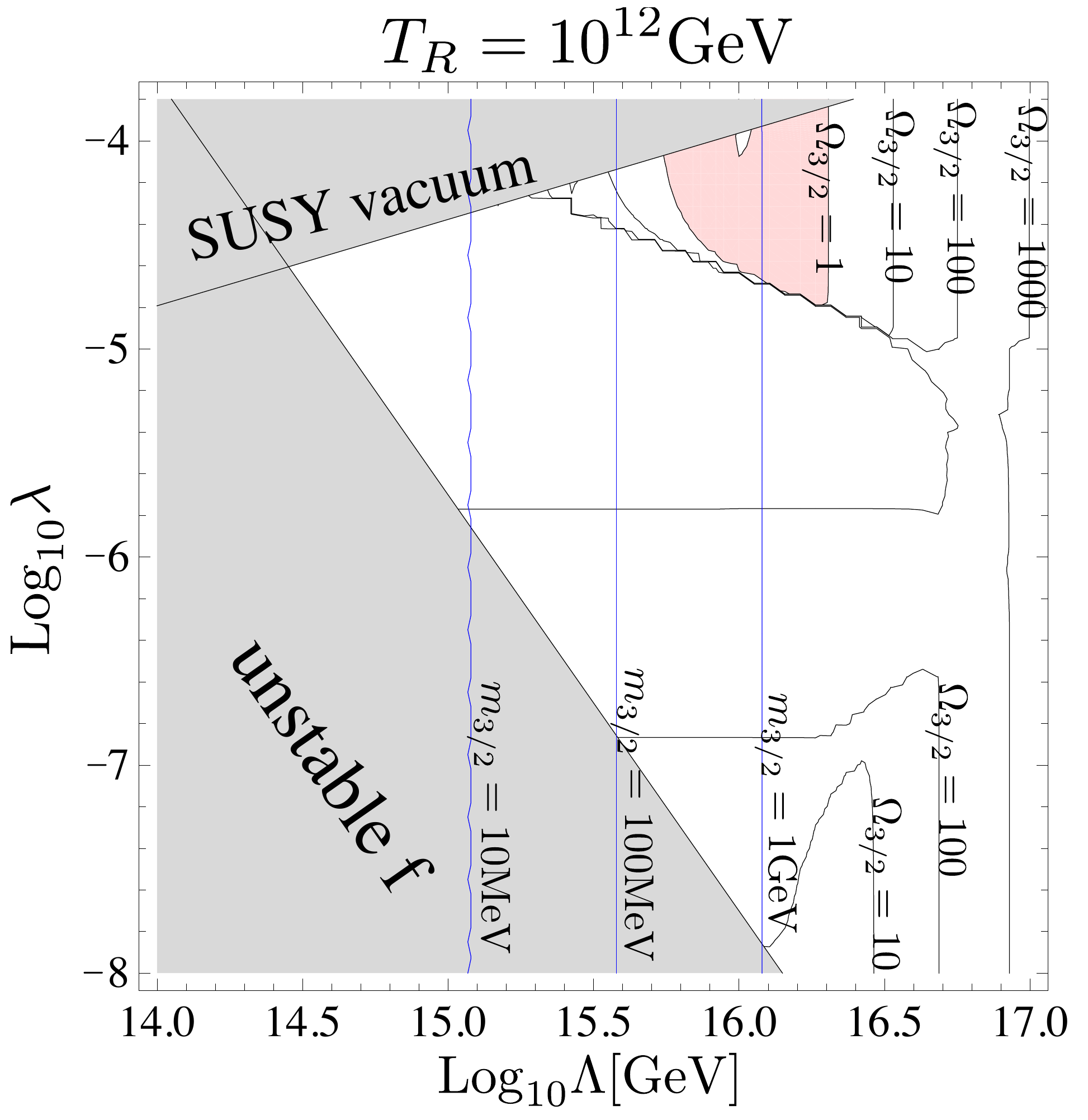} 
\end{center}
\caption{
The contour plot of the total gravitino abundance $\Omega_{3/2}^{\rm tot}$.
Red regions correspond to DM consistent regions $0.1 \leq \Omega_{3/2} \leq 1$.
In $T_R = 10^5 {\rm GeV}$ case thermally produced gravitino comes dominantly from the MSSM sector.
The contribution from the messenger sector becomes comparable to that of the MSSM sector in $T_R = 10^7 {\rm GeV}$.
In $T_R = 10^{12} {\rm GeV}$ case the thermal component consists of the messenger sector contribution.
Non-thermal components from the decay of the pseudo-modulus field are also included in the figures.
}
\label{gravitino}
\end{figure}

We show in Fig.~\ref{gravitino} the contours of the total gravitino abundance
for $T_R= 10^5$, $10^7$ and $10^{12} {\rm\, GeV}$.
Considering ${\mathcal O}(1)$ uncertainties in the calculations
 of dilution factor and thermal gravitino productions, we show 
the highlighted region (in red) where the gravitino abundance $0.1 \leq \Omega_{3/2} \leq 1$ is obtained. 
We expect that somewhere in this region will provide the observationally consistent 
DM abundance, $\Omega_{\rm DM} = 0.2$.
For $T_R = 10^5 {\rm \,GeV}$, the relic gravitino is mostly thermally
produced one. 
On the other hand, for $T_R =10^{12} {\rm \,GeV}$, there is an allowed
region where the dilution factor $\Delta$ is about $10^3 - 10^6$. In this region,
the observed DM abundance is a mixture of the thermally and non-thermally
produced gravitinos. It is crucial to take account of the fact that the gravitino
abundance is independent of the reheating temperature as in
Eq.~\eqref{omega32th}. 

We emphasize here that the contours of the gravitino abundance remain
almost intact for $T_R \gtrsim 10^9$\,GeV. This is because the gravitino production
rate is modified at a temperature above the messenger scale. This has a crucial 
impact on the leptogenesis as we shall see below.

The gravitino mass in the shaded region ranges from $10$\,MeV to a few GeV
(See Eq.~(\ref{m32})). The thermally produced gravitinos therefore behave as
cold dark matter. On the other hand, the non-thermally produced gravitinos 
have a relatively large velocity, which results in a free streaming length of 
${\cal O}(10-100)$\,kpc depending on the parameters~\cite{Ibe:2006rc,Hamaguchi:2009hy}.
In our scenario, the DM consists of a mixture of thermally and non-thermally 
produced gravitinos for $T_R \gtrsim \GEV{7}$, and such partial suppression 
of the density perturbation at small scales may have an interesting impact on the
large scale structure of the Universe.

\subsection{Leptogenesis}
Interestingly, in this model, it is possible to create the right amount
of the baryon asymmetry through the thermal leptogenesis~\cite{Fukugita:1986hr} while satisfying
$\Omega_{3/2} \lesssim 0.2$.

In the thermal leptogenesis, there is an upper bound on the baryon
asymmetry for a fixed reheating temperature~\cite{Davidson:2002qv, Giudice:2003jh, Buchmuller:2005eh}:
\begin{align}
\eta_B \lesssim  5 \times 10^{-11}  \left( \frac{T_R}{10^9 {\rm GeV}} \right) .
\end{align}
This often causes a tension with the gravitino overproduction.
In the presence of the late-time entropy production, one can increase the reheating
temperature because the thermally produced gravitinos are diluted .
However, this does not solve the tension because both the baryon asymmetry
and the conventional formula of the graviton abundance (\ref{omega32usual})
are proportional to $T_R$.

In our scenario, the reheating temperature can be much higher than
$10^9$~GeV without having a problem of gravitino overproduction, 
because the gravitino abundance becomes independent of the reheating temperature
for sufficiently high $T_R$. Recall that the contours of the gravitino abundance in Fig.~\ref{gravitino}
are almost the same for $T_R \gtrsim 10^9$\,GeV.

By comparing Fig.~\ref{dilution} with Fig.~\ref{gravitino}, we see that
the size of the dilution factor $\Delta$ is about $\sim 10^3 - 10^6$ in the
region where $\Omega_{3/2} \simeq 0.1 - 1$.  
Therefore, if the reheating temperature is higher than $\sim 10^{12}
{\rm GeV}$, even though it is partially diluted, the observed amount of
baryon asymmetry can be generated by thermal leptogenesis. This is one of the distinctive features of
our scenario.

\section{Summary}
We have followed the cosmological evolutions of the SUSY breaking pseudo-modulus
field and the messenger fields in a simple gauge mediation model which
break SUSY at a meta-stable state.  We adopt an initial condition such that the pseudo-modulus
field is close to the origin, stabilized by thermal potential via interactions with the messenger
fields in thermal plasma. Such an initial condition can be naturally realized if the U(1)$_R$ symmetry
remains a good symmetry during inflation and $S$ is stabilized near the origin due to 
the positive Hubble-induced mass. We have found that this simple gauge mediation model
with the initial condition is cosmologically viable in a sense that the pseudo-modulus can
settle down at the correct SUSY breaking minimum and that the gravitino relic abundance can
explain the DM of the Universe. Furthermore, thermal leptogenesis is possible without the gravitino
overproduction.

We have numerically calculated the total gravitino relic
abundance, both the thermally produced one taking account of the entropy dilution and the
non-thermally produced one,  and shown that the observed DM density can be explained
for $m_{3/2} = \mathcal{O} (10) {\rm MeV} - {\cal O}(1) {\rm GeV}$.

The messenger fields play a crucial role in the scenario.  
The messengers acquire a thermal mass when the pseudo-modulus stays near the origin, 
which prevents the messengers to fall into the SUSY preserving minimum.  
At the reheating, the gravitino production rate is modified
if the messenger fields are in the thermal plasma. This has a great impact on the gravitino
abundance; it becomes independent of the reheating temperature of the Universe. 
It is this fact that enables thermal leptogenesis to create the right amount of baryon asymmetry
without overproduction of gravitinos.

\section*{Acknowledgments}
This work was supported by the Grant-in-Aid for Scientific Research
on Innovative Areas (No.24111702, No.21111006 and No.23104008) [FT],
Scientific Research (A) (No.22244030 and No.21244033 [FT]), and JSPS
Grant-in-Aid for Young Scientists (B) (No.23740165 [RK] and
No.24740135 [FT]).  This work
was also supported by World Premier International Center Initiative
(WPI Program), MEXT, Japan.

\appendix
\section{Pseudo-modulus potential at a finite temperature}
\label{app:thermal}
The finite temperature effective potential up to one-loop is given by\cite{Dolan:1973qd,Quiros:1999jp}
\begin{align}
V=V_{\rm tree} + V_1 + V_{\rm thermal},
\end{align}
where $V_{\rm tree}$ is the classical potential calculated from Eq.(\ref{kaler}) and (\ref{super}) and $V_1$ is
the zero-temperature one-loop potential calculated from Eq.(\ref{kaler1}).
Finite temperature one-loop correction is 
\begin{multline}
V_{\rm thermal} = \frac{T^4}{2 \pi^2}
\left[
\int_0^{\infty} dx x^2 \sum_i {\log} [1-e^{-\ro{x^2 + (M_S^2)_i / T^2}}] \right. \\
\left. -2\int_0^{\infty} dx x^2 \sum_r {\log} [1+e^{-\ro{x^2 + (M_F^2)_r / T^2}}]
+3 \int_0^{\infty} dx x^2 \sum_a {\log} [1-e^{-\ro{x^2 + (M_V^2)_a / T^2}}]
\right],
\label{VTfull}
\end{multline}
where the three terms represent the contributions from
real scalar fields $\phi_i$, Weyl fermions $\psi_r$ and vector bosons $A^\mu_a$ with 
the eigenvalues of the squared mass matrices $( M_S^2 )_i$, $(M_F^2)_r$ and $(M_V^2)_a$.
For a high temperature limit of $T \gg M_S, M_F, M_V$, the potential can be expanded as
\begin{multline}
V_{\rm thermal}
= - \frac{\pi^2 T^4}{90} \left( N_B + \frac{7}{8} N_F \right)
+ \frac{T^2}{24} \left[
{\Tr} (M_S^2 ) + 3 {\Tr} (M_V^2 )+ {\Tr} (M_F^2 )
\right] \\
-\frac{T}{12 \pi} \left[
{\Tr} (M_S^3 ) + 3 {\Tr} (M_V^3 ) \right]
+ {\cdots} .
\end{multline}

\end{document}